\documentclass[preprint,showpacs,preprintnumbers,superscriptaddress,amsmath,amssymb]{revtex4}

\usepackage{graphicx}
\usepackage{dcolumn}
\usepackage{bm}

\begin{document}

\title{Repulsive and restoring Casimir forces with left-handed materials}

\author{Yaping Yang}
\affiliation{Department of Physics, Tongji University, Shanghai
200092, China}
\author{Ran Zeng}
\affiliation{Department of Physics,
Tongji University, Shanghai 200092, China} \affiliation{Department
of Physics, Harbin Institute of Technology, Harbin 150001, China}
\author{Shutian Liu}
\affiliation{Department of Physics, Harbin Institute of
Technology, Harbin 150001, China}
\author{Hong Chen}
\affiliation{Department of Physics, Tongji University, Shanghai
200092, China}
\author{Shiyao Zhu}
\affiliation{Department of Physics, Hong Kong Baptist University,
Hong Kong, China}

\date{\today }

\begin{abstract}
We investigate repulsive Casimir force between slabs containing left-handed
materials with controllable electromagnetic properties. The sign of Casimir
force is determined by the electric and magnetic properties of the
materials, and it is shown that the formation of the repulsive force is
related to the wave impedances of two slabs. The sign change of the Casimir
force as a function of the distance is studied. Special emphasis is put on
the restoring Casimir force which may be found to exist between perfectly
conducting material and metamaterial slabs. This restoring force is a
natural power for the system oscillation in vacuum and also can be used for
system stabilization.
\end{abstract}


\maketitle

It is well known that the change of the zero point energy of quantized
electromagnetic field in the presence of the boundary surface gives rise to
forces on macroscopic bodies. The existence of the forces called Casimir
effect has attracted considerable attention over decades \cite{bordag}.
Various calculation techniques have been developed where systems of
different geometries of the boundary were considered, and corrections to the
idealized system were made in any practical situation. In recent years, the
force can be measured in experiment with the development of
microelectromechanical and nanoelectromechanical systems (MEMS and NEMS) and
nanotechnology. Since the attractive interaction of a pair of neutral
perfectly conducting parallel plates placed in the vacuum is theoretically
proposed in 1948 by Casimir \cite{casimir}, the forces have been known to be
attractive in most cases. The attractive Casimir forces could lead to
stiction problem in MEMS and NEMS, and therefore, the repulsive forces may
avoid that limits and are of possible practical significance \cite%
{buks,kenneth}.

Recently, artificial composite metamaterials with controllable
electromagnetic properties, namely left-handed material (LHM) \cite%
{smith,parazzoli,dolling,lezec} and single-negative (SNG) material \cite%
{pendry,pendry2,pendry3}, were fabricated experimentally. The LHM has
simultaneously negative permittivity and permeability over a band of
frequencies and thereby the refractive index is negative, while SNG material
has only one negative material parameter within a certain frequency range,
including the $\epsilon $-negative (ENG) media with $\epsilon <0$ (but $\mu
>0$) and the $\mu $-negative (MNG) media with $\mu <0$ (but $\epsilon >0$).
They may possess noticable magnetic properties. The LHM was originally
predicted in the theoretical point of view by Veselago \cite{veselago} where
an alternative solution to Maxwell's equations is provided. The
electromagnetic wave propagating in an LHM has the wave vector opposite to
the direction of energy flow. The directions of the electric field, the
magnetic field and the wave vector form a left-handed system. Such unusual
phenomena as reverse Doppler effect, Cherenkov radiation, anomalous
refraction and reversal of radiation pressure, are expected in the LHM. More
possible applications, such as perfect lens \cite{pendry4}, indirect quantum
interference \cite{yang1}, have been proposed. Recently the Casimir effect
between LHM slabs has been investigated \cite{yang2}. Perfect lens is
introduced in the planar geometry to obtain repulsive Casimir force \cite%
{leonhardt}. SNG materials have also exhibited special features. For
example, unique transmission properties have been found in SNG multilayers
\cite{fredkin,alu}.

In the present work, we analyze the repulsive Casimir effect between two
parallel slabs and the sign change of the force with respect to the slab
separation. The metamaterials are taken into consideration. It is known that
the repulsive Casimir force is possible between two bodies that possess an
asymmetric nature \cite{kenneth2}. The difference of the electromagnetic
properties between two materials plays an important role in forming the
repulsive force. Starting from the relationship between the wave impedances
of two materials and the condition for the formation of the repulsive
Casimir force, we study the different configurations of parallel slabs,
including the ordinary dielectric materials, the metamaterials, and even the
perfectly conducting materials. One may adjust the characteristic
frequencies of the metamaterials in order that the repulsive forces can be
found. In particular, we focus on a special case of sign change of the
force: the repulsive force becomes attractive with the increasing slab
separation. This kind of restoring force, which may exists between a
perfectly conducting slab and a metamaterial slab, could make the slabs
oscillating, or stabilize the two slabs at a fixed distance.

Let us consider the configuration with two parallel infinite slabs, A and B,
each of thickness $d$ separated by a distance $a$, in free space. Based on
the stress tensor method using the properties of the macroscopic field
operators \cite{tomas}, we can calculate the Casimir force, which is
eventually expressed as

\begin{equation}
F_{C}=\frac{\hbar }{2\pi ^{2}}\int_{0}^{\infty }d\xi \int_{0}^{\infty }kdk%
\sqrt{\frac{\xi ^{2}}{c^{2}}+k^{2}}\sum_{N=\text{TE,TM}}\frac{r_{N}^{A}(\xi
,k)r_{N}^{B}(\xi ,k)e^{-2a\sqrt{\xi ^{2}/c^{2}+k^{2}}}}{1-r_{N}^{A}(\xi
,k)r_{N}^{B}(\xi ,k)e^{-2a\sqrt{\xi ^{2}/c^{2}+k^{2}}}}  \label{e01}
\end{equation}%
where $r_{N}^{A(B)}$ is the slab reflection coefficient for TE- and
TM-polarized waves. It is seen from Eq.~(\ref{e01}) that the integrand can
be negative only if the reflection coefficients of the two slabs, $%
r_{N}^{A}(\xi ,k)$ and $r_{N}^{B}(\xi ,k)$, have different signs, which
contributes to the formation of the repulsive force. Therefore, the
repulsive Casimir effect is to be expected when the two parallel slabs have
different electromagnetic properties, and that is to say, the force between
two identical slabs must be attractive, just as has been recently proved in
Ref.~\cite{kenneth2}. The van der Waals force between a magnetically
polarizable particle and a electrically polarizable particle is repulsive
\cite{boyer}, and similarly, repulsive Casimir force is found between a
perfectly conducting plate and an infinitely permeable plate \cite{schaden},
which can be easily calculated from Eq.~(\ref{e01}) by considering $\epsilon
_{A}\rightarrow \infty $ and $\mu _{B}\rightarrow \infty $: $F_{C}=-7\hbar
c\pi ^{2}/1920a^{4}=-\frac{7}{8}F_{0}$, where $F_{0}=\hbar c\pi
^{2}/240a^{4} $ is the well-known formula for the attractive Casimir force
between two perfectly conducting plates. A further fact is known that the
forces are possibly repulsive between two molecules when the medium they
immersed in has the intermediate properties between the properties of two
polarizable molecules \cite{israelachvili}. We thus can apply an analogous
analysis for the system of two parallel slabs in vacuum: The repulsive
behavior may possibly appear when the wave impedances $Z_{A(B)}=\sqrt{\mu
_{A(B)}/\epsilon _{A(B)}}$ of two slabs, which are used to demonstrate the
difference of electric and magnetic properties between two slabs, are
smaller and larger than the impedance of vacuum, respectively.

We consider the effect for the semi-infinitely thick slabs, which is an
approximate model following the condition of $a\ll d$, and the slab
reflection coefficients are simplified to the single interface reflection
coefficients. We model the dispersive material constants by the
single-resonance Drude-Lorentz type
\begin{equation}
\left\{ \epsilon ,\mu \right\} =1+\frac{\omega _{P\nu }^{2}}{\omega _{T\nu
}^{2}-\omega ^{2}-i\gamma _{\nu }\omega },  \label{e02}
\end{equation}%
($\nu =e,m$, which refer to $\epsilon $ and $\mu $, respectively) where $%
\omega _{P\nu }$ is the plasma frequency, $\omega _{T\nu }$ is the resonant
frequency and $\gamma _{\nu }$ is the damping frequency. In the following, $%
\omega _{0}$ denotes a unit of the frequency with which the characteristic
frequencies are scaled, and $\lambda _{0}=2\pi c/\omega _{0}$ is the
corresponding wavelength in the vacuum.

\begin{figure}[tbp]
\includegraphics[width=8cm]{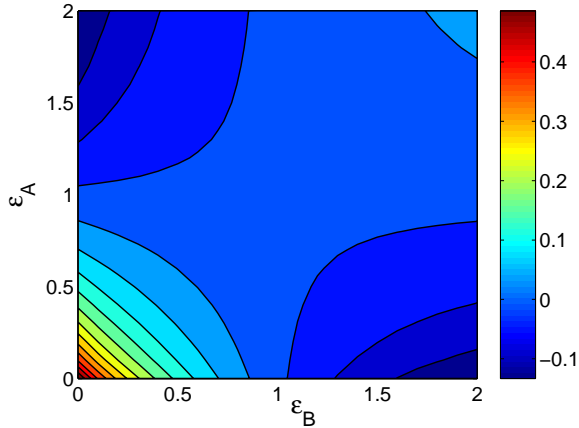}
\caption{Contour plot of $F_r$ between two semi-infinite
non-dispersive
ordinary dielectric slabs as a function of the permittivities $\protect%
\epsilon_A$ and $\protect\epsilon_B$ (slab separation $a=\protect\lambda_0/4$%
).} \label{fig2}
\end{figure}

For the dispersive ordinary dielectric material slabs with trivial constant
permeability $\mu=1$, most values of the impedances of two slabs over the
whole frequency range are no larger than unity, i.e., the impedance of
vacuum. Thus the integrand mainly contributes to the attractive forces, and
it is hard to obtain the repulsive Casimir forces. However, the trend
towards formation of the repulsive forces may be clearly seen from the cases
of the material slabs of constant impedances. The dependence of the relative
force $F_r=F_C/F_0$ between two idealized, non-dispersive ordinary
dielectric slabs on the permittivities is shown in Fig.~\ref{fig2}. As is
expected, the character of the regions where the repulsive forces are found
is that the one of the two slab permittivities is larger than the value for
vacuum, whereas the other is smaller, and moreover, the greater the
difference between two permittivities, the more easily the repulsive force
is obtained. But in practical, it is hard to find real dielectrics with
approximately non-dispersive permittivity smaller than unity over a
sufficiently wide frequency range, and the Casimir forces are generally
attractive between real dispersive ordinary dielectric slabs.

\begin{figure}[tbp]
\includegraphics[width=7.5cm]{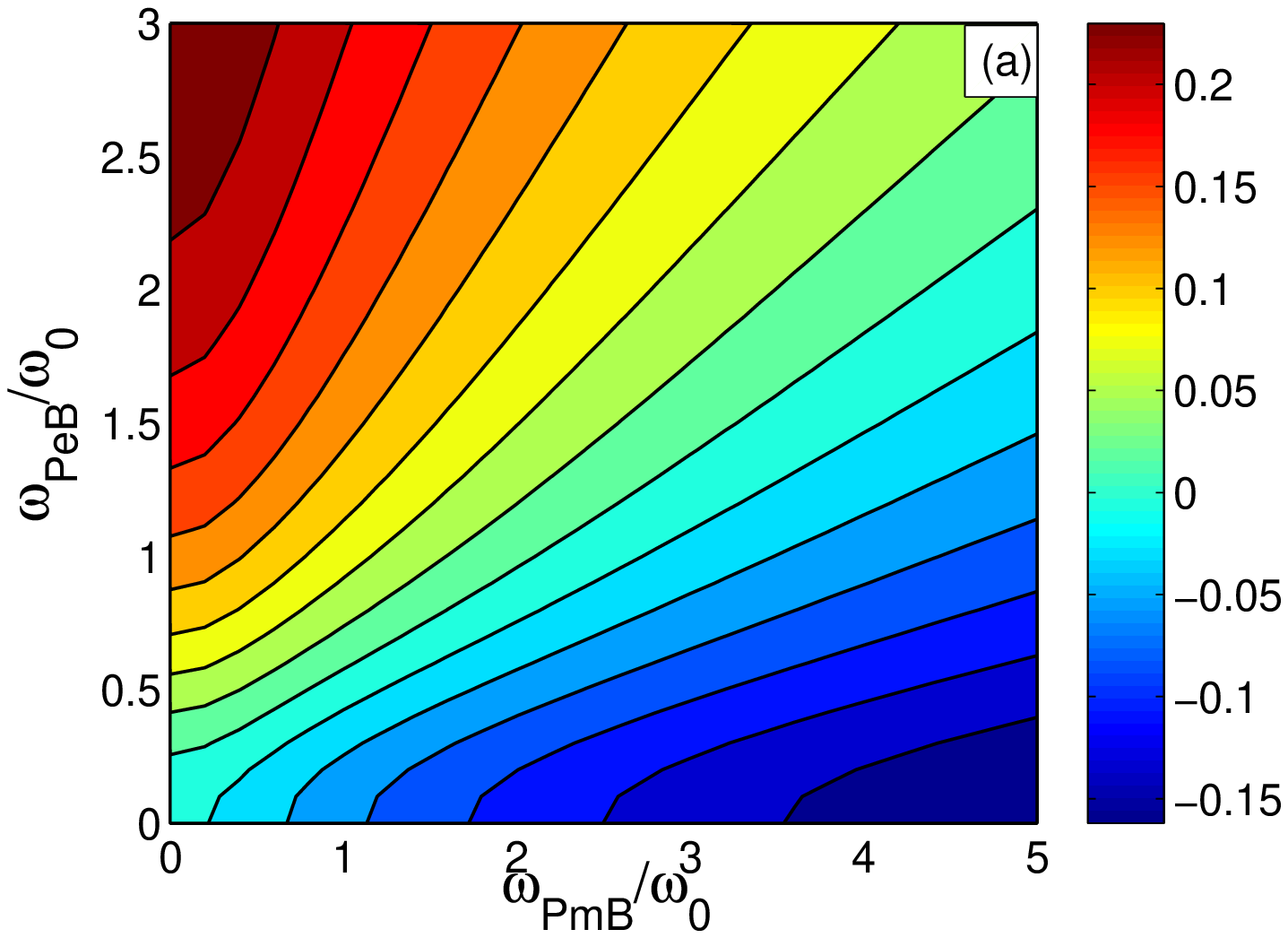}\quad %
\includegraphics[width=7.5cm]{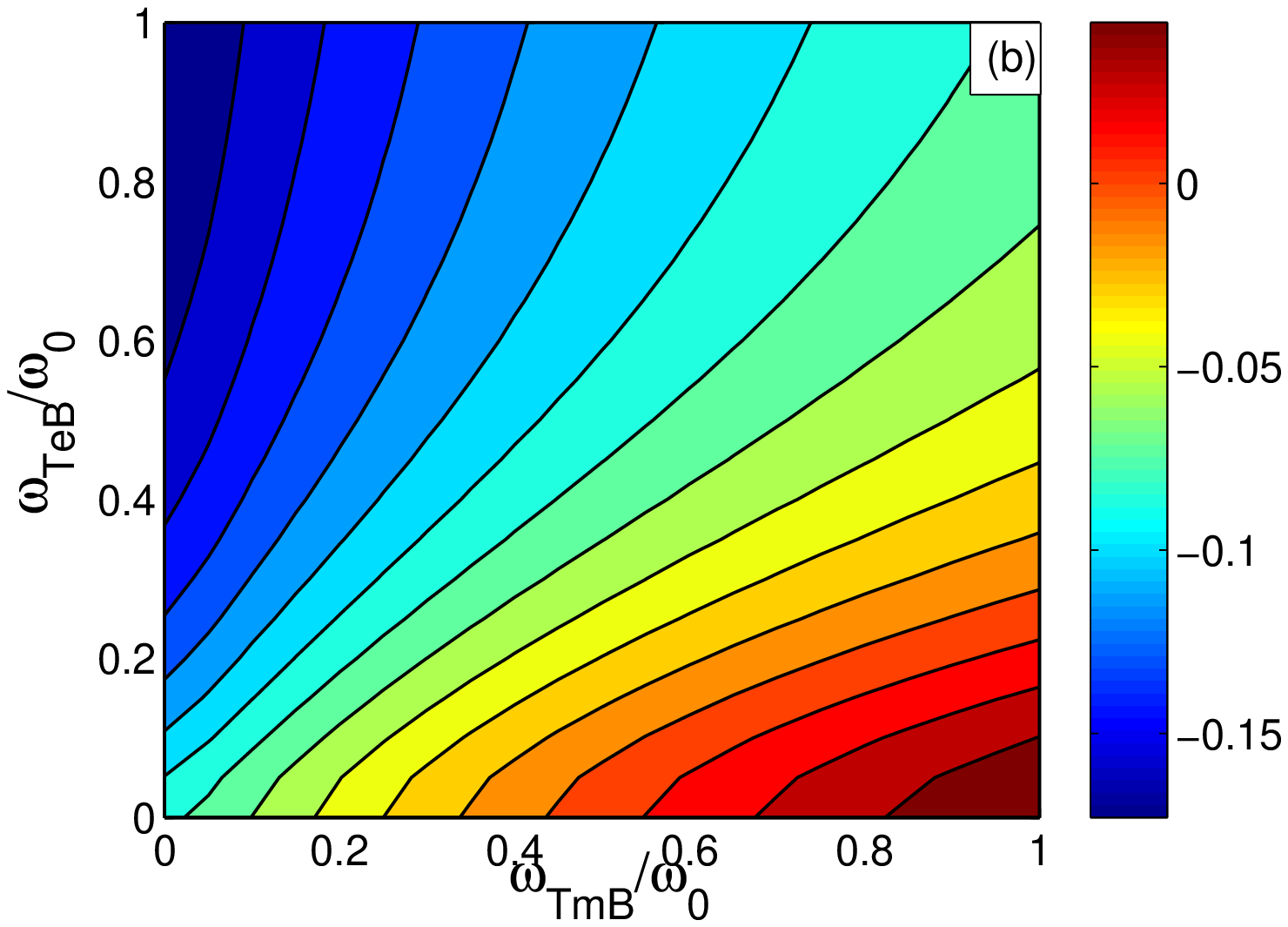}
\caption{Contour plot of $F_r$ between the ordinary dielectric and
metamaterial semi-infinite slabs as a function of the plasma and
the resonant frequencies of the metamaterial [slab
separation $a=\protect\lambda_0/4$; characteristic frequencies: $\protect%
\omega_{PeA}=\protect\omega_0$, $\protect\omega_{TeA}=0$, $\protect\gamma%
_{eA}=10^{-2}\protect\omega_{TeA}$, $\protect\gamma_{\protect\nu B}=10^{-2}\protect%
\omega_{T\protect\nu B}$ ($\protect\nu=e,m$), (a): $\protect\omega_{TeB}=\protect\omega%
_{TmB}=0.5\protect\omega_0$, (b):
$\protect\omega_{PeB}=0.5\protect\omega_0$,
$\protect\omega_{PmB}=3\protect\omega_0$].} \label{fig3}
\end{figure}

It is possible to control the sign of the Casimir force when the
metamaterial with controllable permittivity and permeability described by
Eq.~(\ref{e02}) is introduced. Consider that slab A is the dispersive
ordinary dielectric material and slab B is metamaterial, then we present the
influence of the characteristic frequencies of the metamaterial under the
condition of the parameters of the ordinary dielectric being fixed. Fig.~\ref%
{fig3}(a) illustrates the dependence of the relative Casimir force
on the plasma frequencies of the metamaterial. The repulsive force
tends to appear in the region where $\omega _{PeB}$ is decreased
and $\omega _{PmB}$ is simultaneously increased. This behavior can
be explained as follows. For slab A, most values of the impedance
over the whole frequency range are no larger than unity, that is,
the permittivity is no smaller than the permeability. Therefore,
in order to satisfy the condition of formation of the repulsive
force, slab B must have the permeability that exceeds the
permittivity. We emphasis that repulsive force can be obtained if
$Z_{A}$ and $Z_{B}$ are, respectively, smaller and larger than
unity and the difference between them is great. The decreasing
$\omega _{PeB}$ and the increasing $\omega _{PmB}$ correspond to
the decreasing permittivity and the increasing permeability, and
accordingly the repulsive force tends to appear.
Fig.~\ref{fig3}(b) shows the influence of the resonant frequencies
of the metamaterial. By simultaneously increasing $\omega _{TeB}$
and decreasing $\omega _{TmB}$, the attractive force can be
changed to the repulsive force, which may be explained through a
similar analysis.

\begin{figure}[tbp]
\includegraphics[width=7.5cm]{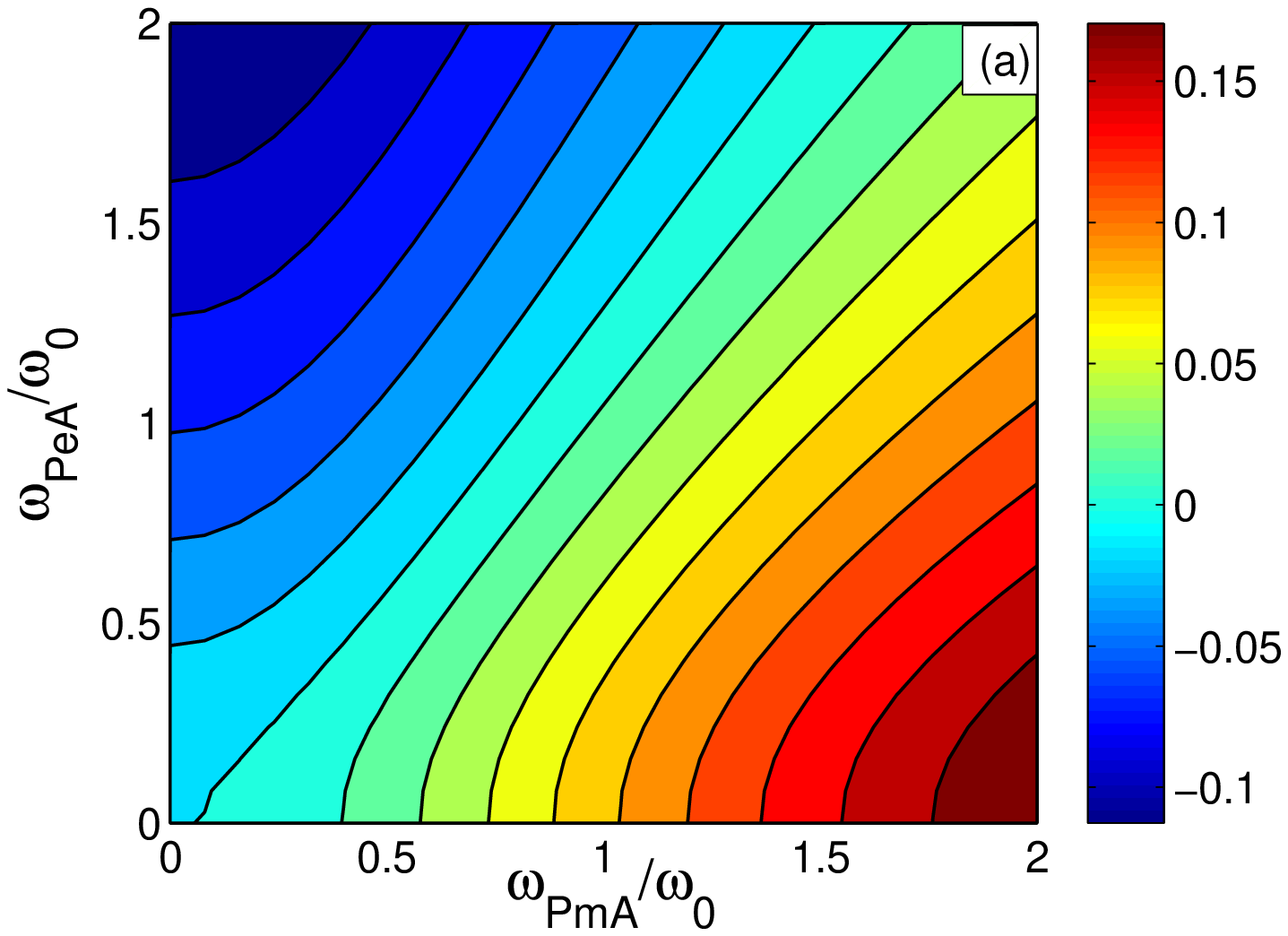}\quad %
\includegraphics[width=7.5cm]{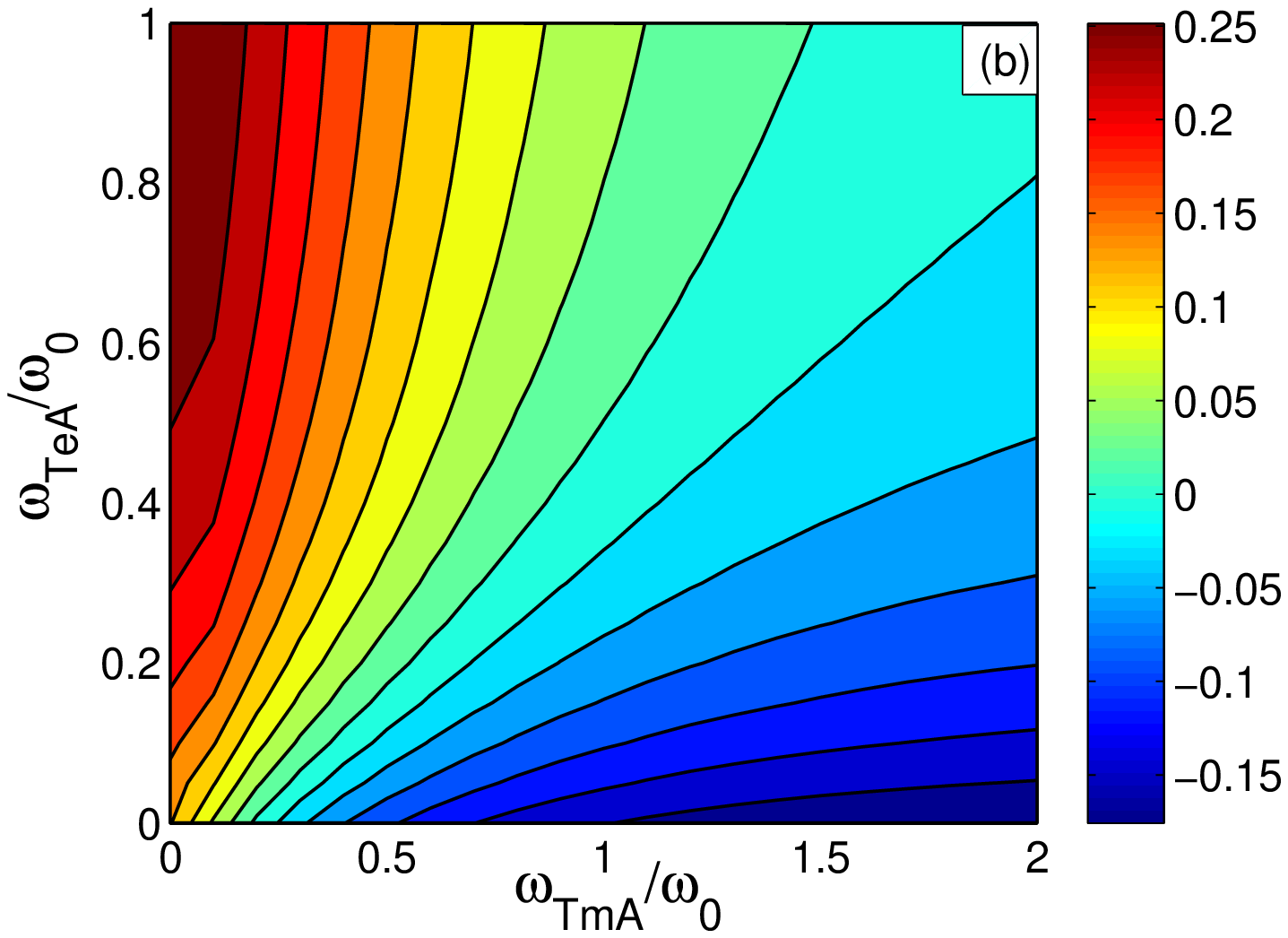}
\caption{Contour plot of $F_r$ between two metamaterial
semi-infinite slabs
as a function of the plasma frequencies $\protect\omega_{PeA}$ and $\protect%
\omega_{PmA}$ [slab separation $a=\protect\lambda_0/4$; characteristic
frequencies: $\protect\omega_{TeB}=0.7\protect\omega_0$, $%
\protect\omega_{TmB}=0.5\protect\omega_0$, $\protect\omega_{PeB}=0.2\protect%
\omega_0$, $\protect\omega_{PmB}=1.5\protect\omega_0$, $\protect\gamma_{%
\protect\nu \protect\zeta}=10^{-2}\protect\omega_{T\protect\nu \protect\zeta%
} $ ($\protect\nu=e,m$, $\protect\zeta=A,B$), (a): $\protect\omega_{TeA}=0.5\protect\omega_0$, $\protect\omega%
_{TmA}=\protect\omega_0$, (b):
$\protect\omega_{PeA}=\protect\omega_{PmA}=\protect\omega_0$].}
\label{fig5}
\end{figure}

It makes the sign control more freely when both the slabs are the
metamaterials. Then we consider the influences of the
characteristic frequencies of one metamaterial (slab B) when the
parameters of the other metamaterial (slab A) are fixed, which are
shown in Fig.~\ref{fig5}. In order to distinguish from the case of
dispersive ordinary dielectric slab, we let slab B have the
magnetic properties stronger than the electric properties. As the
parameters of slab B chosen in the figures indicate, the magnetic
resonant frequency is lower than the electric one, and the
magnetic plasma frequency higher than the electric one, thus the
impedances are in general larger than the impedance of vacuum. The
appearance and gradually increase in magnitude of repulsive forces
correspond to simultaneously stronger electric properties and
weaker magnetic properties, which is opposite to the cases shown
in Fig.~\ref{fig3}. This confirms again the condition of the
formation of the repulsive Casimir force that is stated above.

%

\begin{figure}[tbp]
\includegraphics[width=8cm]{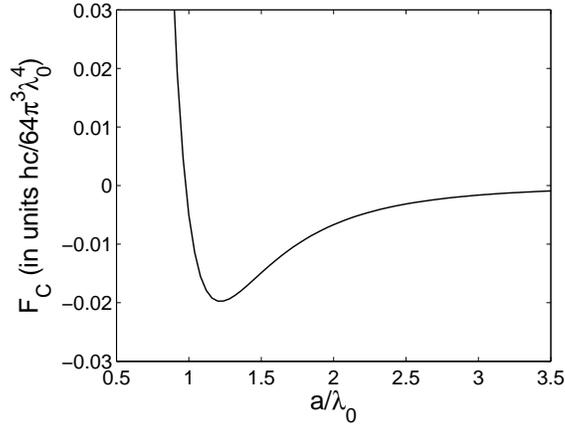}
\caption{Casimir force $F_C$ between two metamaterial semi-infinite slabs as
a function of the slab separation $a$ [characteristic frequencies: $\protect%
\omega_{PeA}=\protect\omega_0$, $\protect\omega_{TeA}=0.5\protect\omega_0$, $%
\protect\omega_{PmA}=\protect\omega_{TmA}=\protect\omega_0$, $\protect\omega%
_{PeB}=0.2\protect\omega_0$, $\protect\omega_{TeB}=0.7\protect\omega_0$, $%
\protect\omega_{PmB}=1.5\protect\omega_0$, $\protect\omega_{TmB}=0.5\protect%
\omega_0$, $\protect\gamma_{\protect\nu \protect\zeta}=10^{-2}\protect\omega%
_{T\protect\nu \protect\zeta}$ ($\protect\nu=e,m$, $\protect\zeta=A,B$)].}
\label{fig8}
\end{figure}

We proceed now to the dependence of the force on the distance between the
slabs. The existence of the boundary surface in vacuum changes the
zero-point energy, thus the mode densities are redistributed. There are more
modes for certain frequencies between the macroscopic slabs than in the
space outside the slabs, whereas there are fewer for other frequencies \cite%
{hushwater}. Consequently the force resulted from the inside and outside
pressure difference can be attractive or repulsive. When the slab separation
is adiabatically changed, the redistribution of the mode densities are
correspondingly influenced, and then the Casimir force may be shifted to
opposite direction. One can adjust the characteristic frequencies of the
metamaterials to satisfy the condition of formation of repulsive forces that
has been stated above at certain slab separations. In general, the Casimir
force is attractive at very short distances, and as the slabs get further
away from each other, the attractive force is gradually lowered and may
becomes repulsive at certain distance, as is seen from Fig.~\ref{fig8}. The
repulsive force begins to grow larger and attains a maximum magnitude value,
and then decreases again with the increasing slab separation.

\begin{figure}[tbp]
\includegraphics[width=7.5cm]{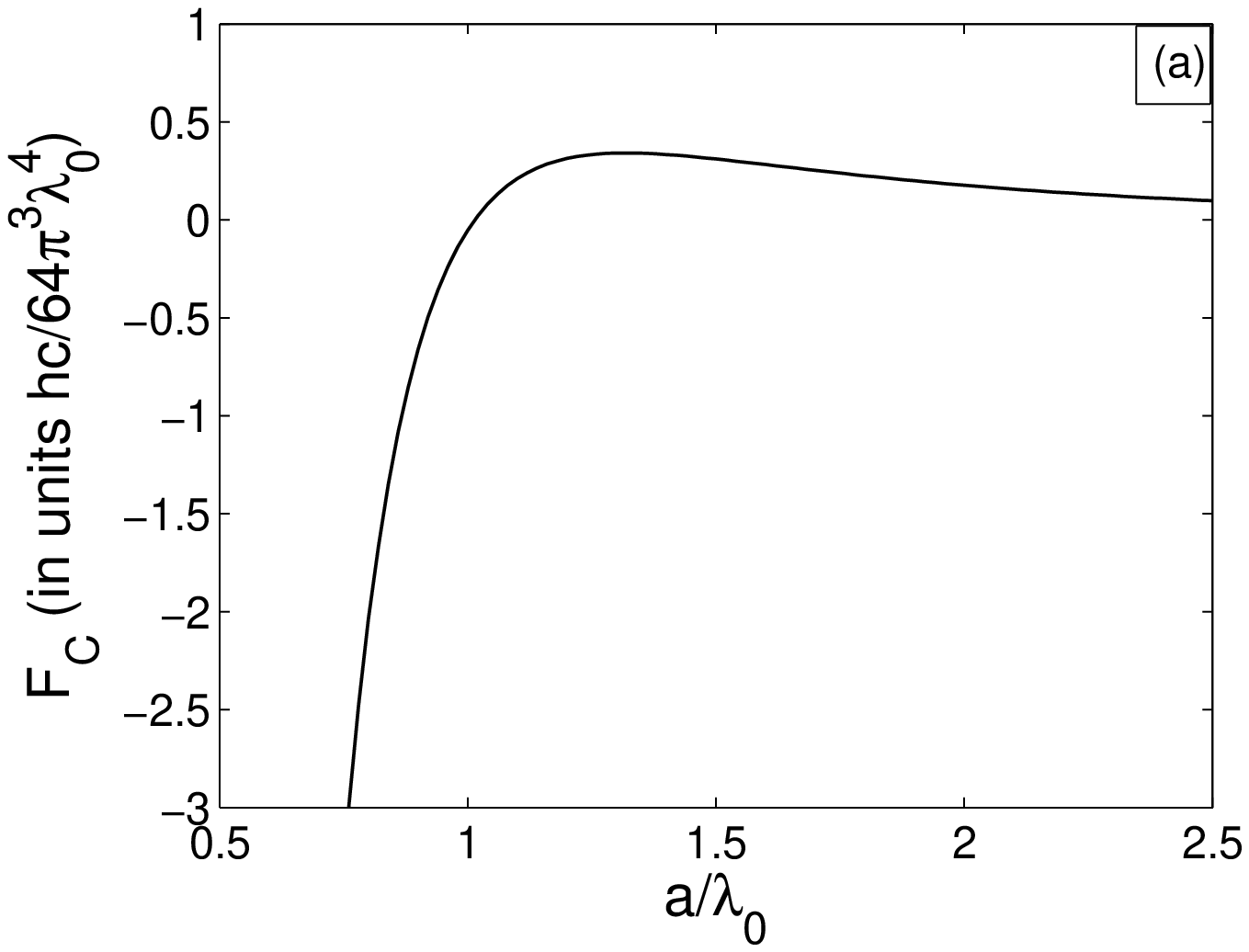}\quad %
\includegraphics[width=7.5cm]{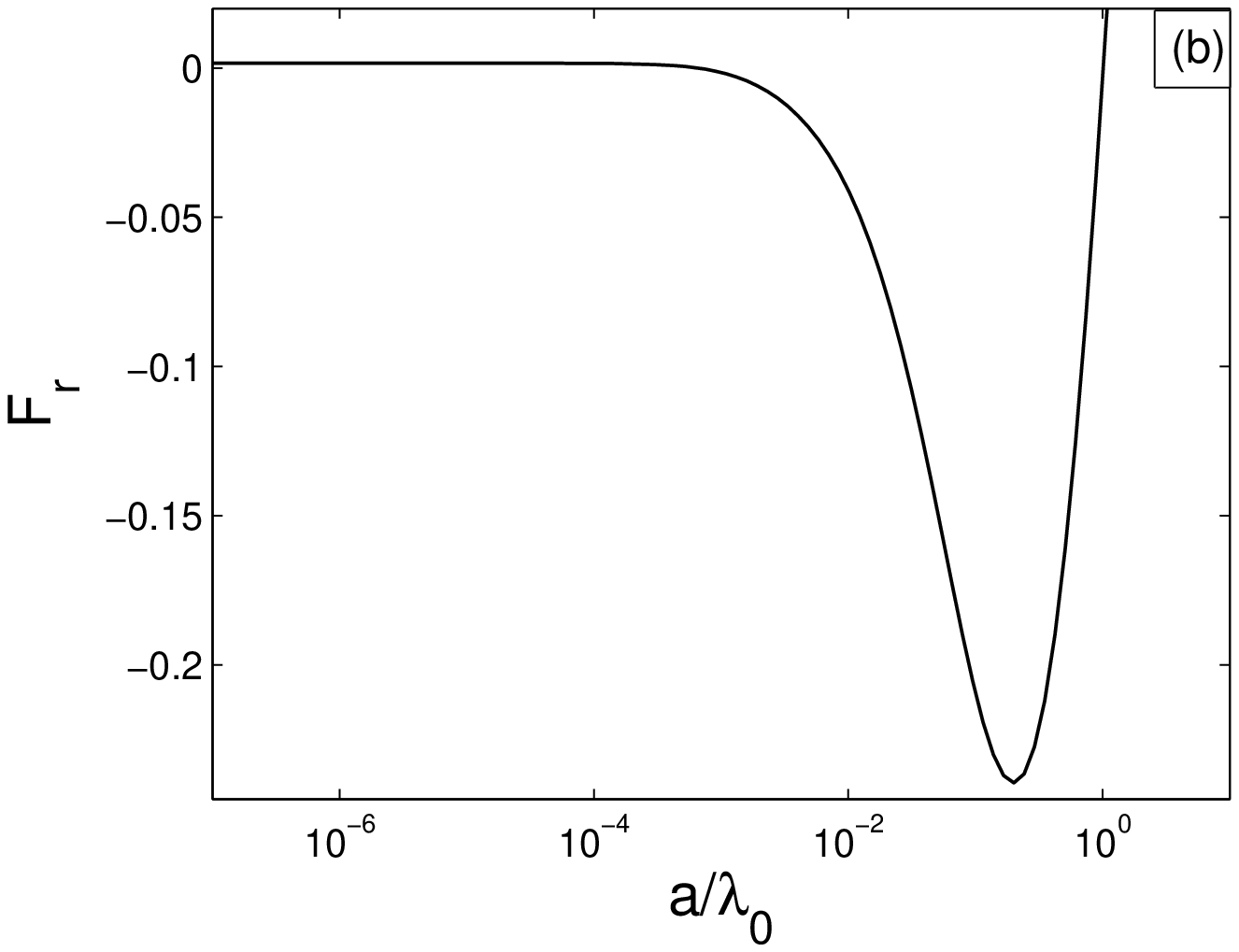}
\caption{Casimir force between a perfectly conducting
semi-infinite slab and a metamaterial semi-infinite slab as a
function of the slab
separation. (a) and (b) show the dependence within different distance regions. [characteristic frequencies of the metamaterial: $\protect%
\omega_{PeB}=0.5\protect\omega_0$, $\protect\omega_{TeB}=10^{-3}\protect%
\omega_0$, $\protect\omega_{PmB}=3\protect\omega_0$, $\protect\omega%
_{TmB}=0.7\protect\omega_0$, $\protect\gamma_{\protect\nu B}=10^{-2}\protect%
\omega_{T\protect\nu B}$ ($\protect\nu=e,m$)].}
\label{fig9}
\end{figure}

An opposite situation can be found if one of the slabs is taken to be
perfect conductor, which has the infinite permittivity $\epsilon \to \infty$
and accordingly the properties of perfect reflections $r_{TE}=-1$ and $%
r_{TM}=1$. Through the tuning of the parameters of the other slab
(metamaterial), one can obtain the reversion of the Casimir force
sign change. As Fig.~\ref{fig9}(a) indicated, the Casimir force
between the perfectly conducting and metamaterial slabs may change
from repulsion to attraction with the increasing slab separation.
The existence of this type of restoring force presents the
possibilities of the quasi-harmonic oscillation for the mechanical
system in vacuum, and otherwise it may stabilize the system. Then
we investigate the formation of this restoring force.

\begin{figure}[tbp]
\includegraphics[width=7.5cm]{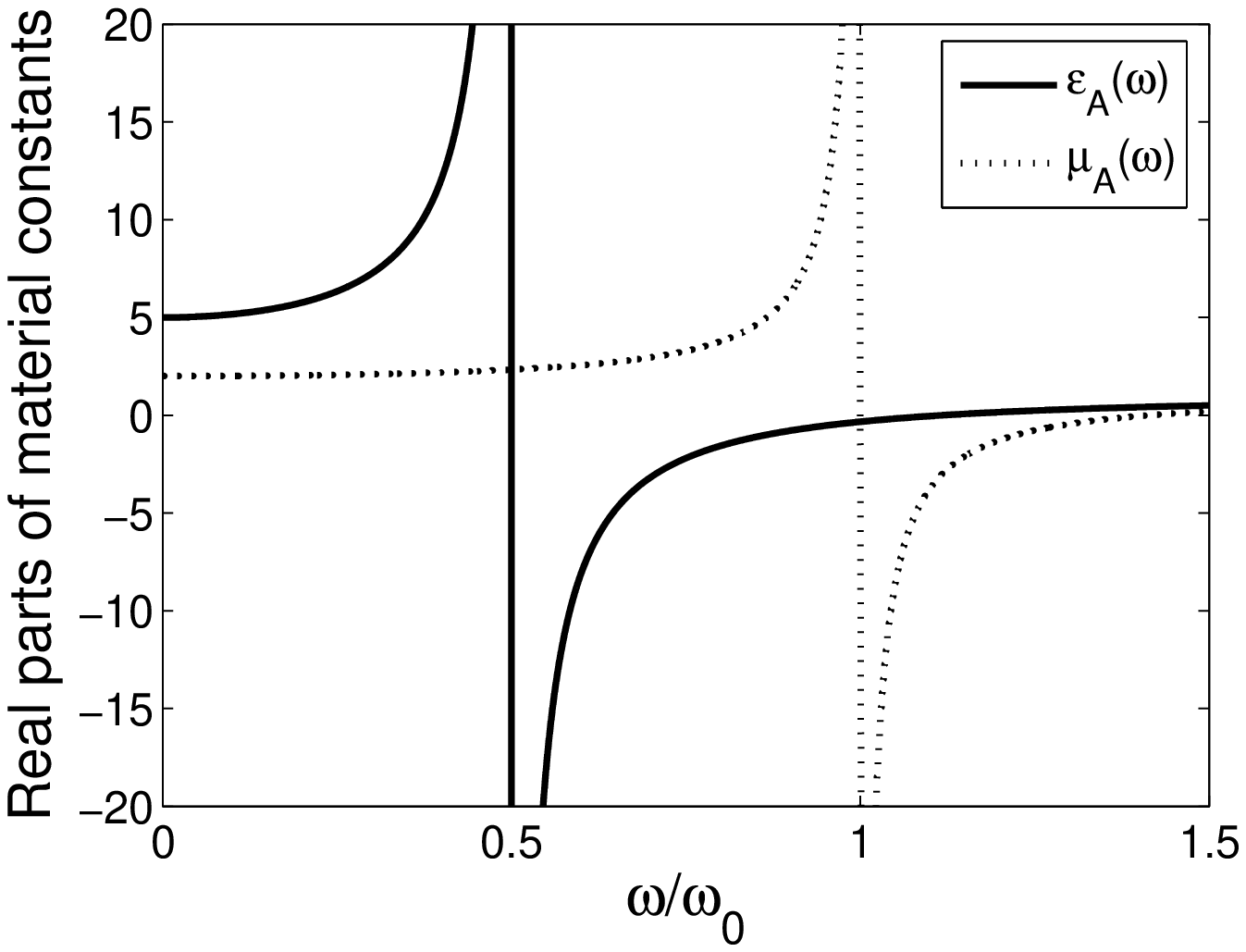}\quad %
\includegraphics[width=7.5cm]{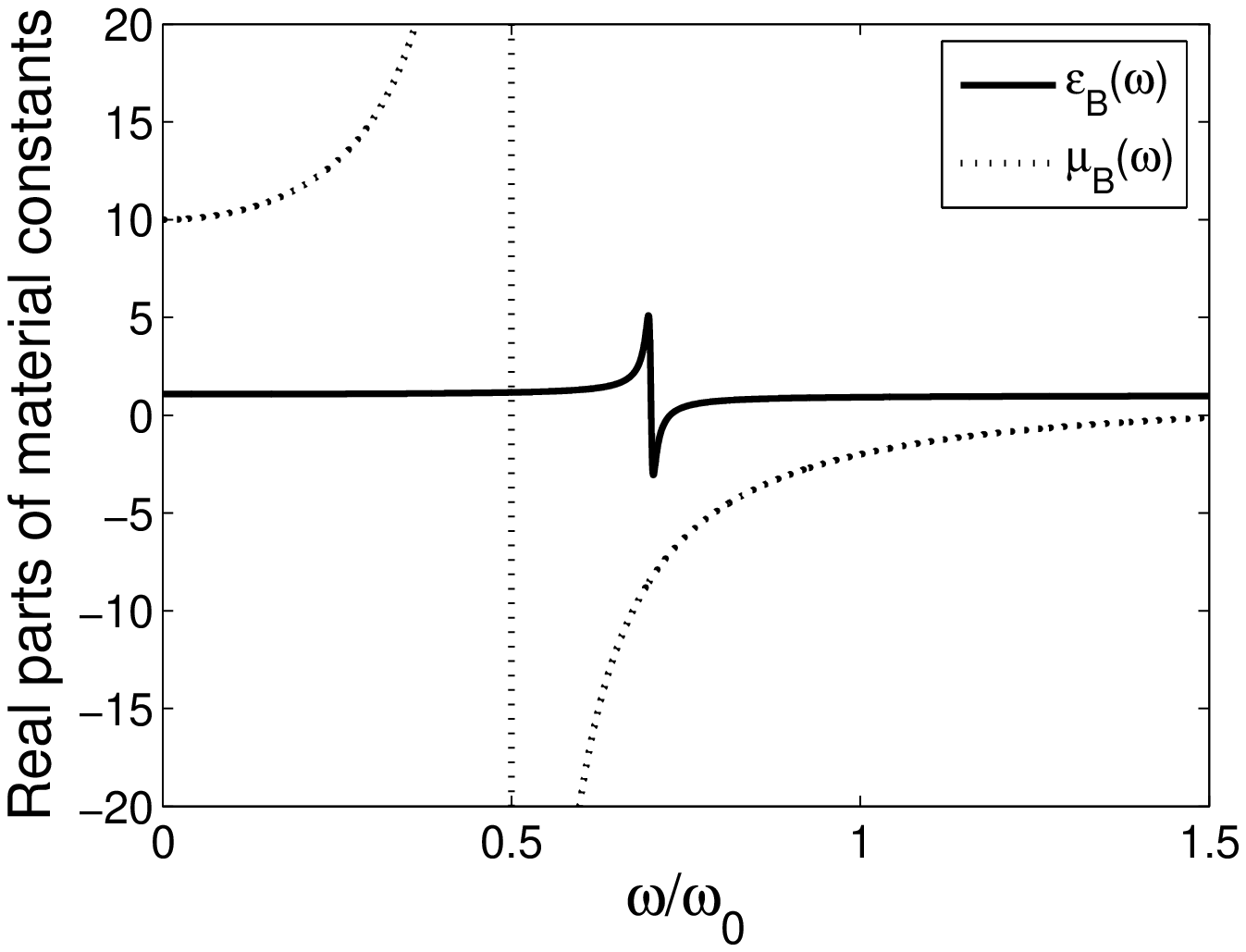}
\caption{Real part of the $\protect\epsilon$ and $\protect\mu$ of two
metamaterials A and B chosen in Fig.~\protect\ref{fig8} versus frequency.}
\label{fig10}
\end{figure}

\begin{figure}[tbp]
\includegraphics[width=8cm]{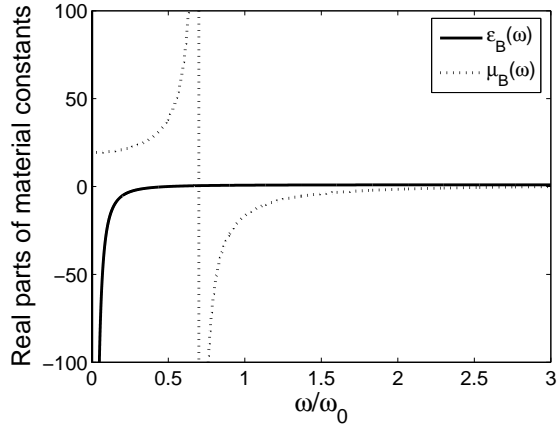}
\caption{Real part of the $\protect\epsilon$ and $\protect\mu$ of the
metamaterial chosen in Fig.~\protect\ref{fig9} versus frequency.}
\label{fig11}
\end{figure}

When the distances are large, the corresponding effective frequency range
that has main contribution to the Casimir force integral is narrowed down to
low frequencies. For the case of two metamaterial semi-infinite slabs as
shown in Fig.~\ref{fig8}, the static electromagnetic properties are much
different between two slabs (cf. Fig.~\ref{fig10}). The static impedances of
two metamaterials are evaluated as: $Z_A(0)=\sqrt{\mu_A(0)/\epsilon_A(0)}%
\simeq 0.63$, $Z_B(0)=\sqrt{\mu_B(0)/\epsilon_B(0)}\simeq 3.04$. Thus in the
large separation region the force appears to be repulsive. For the case of a
perfectly conducting slab and a metamaterial slab as shown in Fig.~\ref{fig9}(a)%
, the chosen metamaterial slab possesses the static electric properties that
exceed the static magnetic properties (cf. Fig.~\ref{fig11}), which is seen
from its static impedances $Z_B(0)=\sqrt{\mu_B(0)/\epsilon_B(0)}\to 0$. The
perfectly conducting slab, whose impedance $Z\to 0$ at any frequencies, has
the similar electromagnetic properties as the chosen metamaterial slab
within low frequency range, therefore the force between them is attractive
at large distances. When the separation of two slabs is relatively small,
the corresponding effective frequency range is rather wide. For the
metamaterial slab, although the material constants are dispersive, the total
effect may lead to that the magnetic properties are stronger than the
electric properties, or the other way round. Here the three examples of the
metamaterials (slabs A and B in Fig.~\ref{fig10} and slab B in Fig.~\ref%
{fig11}) may all belong to the former situation, thus at short distances,
the attraction results from the similarity of the slabs, and the great
difference between the properties of the perfectly conductor and the
metamaterial gives rise to a repulsive force. The restoring force requests
that there is repulsion at short distances and attraction at large
distances, so in summary there is restoring force between the perfectly
conductor and the metamaterial slabs, whereas the restoring force does not
appear in the case of two metamaterial slabs.

One may adjust the values of the characteristic frequencies in
order to obtain the large-separation attractive force between two
metamaterial slabs. However, by doing so, the force for the
relatively short separation may generally be attractive as well,
accordingly there is yet no restoring force between such
metamaterial slabs. We attribute the formation of the restoring
force to the speciality of the perfect conductor. The perfect
conductor has the extremely strong electric properties at any
frequencies. This extremeness makes the force sign change rather
sensitive with respect to the variation of the properties of the
other slab. The Casimir force for the above case within wider
distance range is plotted in Fig.~\ref{fig9}(b). In fact, the
force is attractive at much shorter distances where the
corresponding effective frequency ranges are quite wider. Thus for
this case, the sign of the casimir force has changed twice with
the increasing slab separation. The restoring force is only
possible under such situation. The forces between perfectly
conducting and metamaterial slabs at extremely short distances may
either be repulsive, but then the properties of the chosen
metamaterial, which give rise to repulsion at those short
distances, generally also lead to repulsion at larger distances,
and there is no restoring force at any distances.

In conclusion, we have studied cases of different directions of
the Casimir force between parallel slabs, including ordinary
dielectrics, metamaterials and perfect conductor \cite{rosa}. It
is found that the repulsive Casimir force can be obtained when the
electromagnetic properties of two slabs are greatly different. We
particularly focus on the restoring Casimir force which is found
between the perfect conductor and the metamaterial, and the
dependence of Casimir force on the slab separation is investigated
from the point of view of the slab electromagnetic properties
within different frequency range.

\bigskip

This work was partly supported by the National Natural Science
Foundation of China (Grant No. 10674103, No. 10634050), the
Foundation of Ministry of Education of China (Grant No.
NCET-06-0384) and the National Basic Research Program of China
(Grant No. 2006CB302901, No. 2006CB921701, No. 2007CB613201).

\end{document}